\documentclass{article}
\usepackage{amsmath}
\usepackage{amsthm}
\usepackage{array}
\usepackage{booktabs}
\usepackage{graphicx}
\usepackage{wasysym}
\usepackage{tabularx}
\usepackage{threeparttable}
\usepackage[newfloat,draft]{minted}
\usepackage{caption}
\usepackage{parcolumns}

\makeatletter
\def\PYG@reset{\let\PYG@it=\relax \let\PYG@bf=\relax%
    \let\PYG@ul=\relax \let\PYG@tc=\relax%
    \let\PYG@bc=\relax \let\PYG@ff=\relax}
\def\PYG@tok#1{\csname PYG@tok@#1\endcsname}
\def\PYG@toks#1+{\ifx\relax#1\empty\else%
    \PYG@tok{#1}\expandafter\PYG@toks\fi}
\def\PYG@do#1{\PYG@bc{\PYG@tc{\PYG@ul{%
    \PYG@it{\PYG@bf{\PYG@ff{#1}}}}}}}
\def\PYG#1#2{\PYG@reset\PYG@toks#1+\relax+\PYG@do{#2}}

\expandafter\def\csname PYG@tok@gd\endcsname{\def\PYG@tc##1{\textcolor[rgb]{0.63,0.00,0.00}{##1}}}
\expandafter\def\csname PYG@tok@gu\endcsname{\let\PYG@bf=\textbf\def\PYG@tc##1{\textcolor[rgb]{0.50,0.00,0.50}{##1}}}
\expandafter\def\csname PYG@tok@gt\endcsname{\def\PYG@tc##1{\textcolor[rgb]{0.00,0.27,0.87}{##1}}}
\expandafter\def\csname PYG@tok@gs\endcsname{\let\PYG@bf=\textbf}
\expandafter\def\csname PYG@tok@gr\endcsname{\def\PYG@tc##1{\textcolor[rgb]{1.00,0.00,0.00}{##1}}}
\expandafter\def\csname PYG@tok@cm\endcsname{\let\PYG@it=\textit\def\PYG@tc##1{\textcolor[rgb]{0.25,0.50,0.50}{##1}}}
\expandafter\def\csname PYG@tok@vg\endcsname{\def\PYG@tc##1{\textcolor[rgb]{0.10,0.09,0.49}{##1}}}
\expandafter\def\csname PYG@tok@vi\endcsname{\def\PYG@tc##1{\textcolor[rgb]{0.10,0.09,0.49}{##1}}}
\expandafter\def\csname PYG@tok@vm\endcsname{\def\PYG@tc##1{\textcolor[rgb]{0.10,0.09,0.49}{##1}}}
\expandafter\def\csname PYG@tok@mh\endcsname{\def\PYG@tc##1{\textcolor[rgb]{0.40,0.40,0.40}{##1}}}
\expandafter\def\csname PYG@tok@cs\endcsname{\let\PYG@it=\textit\def\PYG@tc##1{\textcolor[rgb]{0.25,0.50,0.50}{##1}}}
\expandafter\def\csname PYG@tok@ge\endcsname{\let\PYG@it=\textit}
\expandafter\def\csname PYG@tok@vc\endcsname{\def\PYG@tc##1{\textcolor[rgb]{0.10,0.09,0.49}{##1}}}
\expandafter\def\csname PYG@tok@il\endcsname{\def\PYG@tc##1{\textcolor[rgb]{0.40,0.40,0.40}{##1}}}
\expandafter\def\csname PYG@tok@go\endcsname{\def\PYG@tc##1{\textcolor[rgb]{0.53,0.53,0.53}{##1}}}
\expandafter\def\csname PYG@tok@cp\endcsname{\def\PYG@tc##1{\textcolor[rgb]{0.74,0.48,0.00}{##1}}}
\expandafter\def\csname PYG@tok@gi\endcsname{\def\PYG@tc##1{\textcolor[rgb]{0.00,0.63,0.00}{##1}}}
\expandafter\def\csname PYG@tok@gh\endcsname{\let\PYG@bf=\textbf\def\PYG@tc##1{\textcolor[rgb]{0.00,0.00,0.50}{##1}}}
\expandafter\def\csname PYG@tok@ni\endcsname{\let\PYG@bf=\textbf\def\PYG@tc##1{\textcolor[rgb]{0.60,0.60,0.60}{##1}}}
\expandafter\def\csname PYG@tok@nl\endcsname{\def\PYG@tc##1{\textcolor[rgb]{0.63,0.63,0.00}{##1}}}
\expandafter\def\csname PYG@tok@nn\endcsname{\let\PYG@bf=\textbf\def\PYG@tc##1{\textcolor[rgb]{0.00,0.00,1.00}{##1}}}
\expandafter\def\csname PYG@tok@no\endcsname{\def\PYG@tc##1{\textcolor[rgb]{0.53,0.00,0.00}{##1}}}
\expandafter\def\csname PYG@tok@na\endcsname{\def\PYG@tc##1{\textcolor[rgb]{0.49,0.56,0.16}{##1}}}
\expandafter\def\csname PYG@tok@nb\endcsname{\def\PYG@tc##1{\textcolor[rgb]{0.00,0.50,0.00}{##1}}}
\expandafter\def\csname PYG@tok@nc\endcsname{\let\PYG@bf=\textbf\def\PYG@tc##1{\textcolor[rgb]{0.00,0.00,1.00}{##1}}}
\expandafter\def\csname PYG@tok@nd\endcsname{\def\PYG@tc##1{\textcolor[rgb]{0.67,0.13,1.00}{##1}}}
\expandafter\def\csname PYG@tok@ne\endcsname{\let\PYG@bf=\textbf\def\PYG@tc##1{\textcolor[rgb]{0.82,0.25,0.23}{##1}}}
\expandafter\def\csname PYG@tok@nf\endcsname{\def\PYG@tc##1{\textcolor[rgb]{0.00,0.00,1.00}{##1}}}
\expandafter\def\csname PYG@tok@si\endcsname{\let\PYG@bf=\textbf\def\PYG@tc##1{\textcolor[rgb]{0.73,0.40,0.53}{##1}}}
\expandafter\def\csname PYG@tok@s2\endcsname{\def\PYG@tc##1{\textcolor[rgb]{0.73,0.13,0.13}{##1}}}
\expandafter\def\csname PYG@tok@nt\endcsname{\let\PYG@bf=\textbf\def\PYG@tc##1{\textcolor[rgb]{0.00,0.50,0.00}{##1}}}
\expandafter\def\csname PYG@tok@nv\endcsname{\def\PYG@tc##1{\textcolor[rgb]{0.10,0.09,0.49}{##1}}}
\expandafter\def\csname PYG@tok@s1\endcsname{\def\PYG@tc##1{\textcolor[rgb]{0.73,0.13,0.13}{##1}}}
\expandafter\def\csname PYG@tok@dl\endcsname{\def\PYG@tc##1{\textcolor[rgb]{0.73,0.13,0.13}{##1}}}
\expandafter\def\csname PYG@tok@ch\endcsname{\let\PYG@it=\textit\def\PYG@tc##1{\textcolor[rgb]{0.25,0.50,0.50}{##1}}}
\expandafter\def\csname PYG@tok@m\endcsname{\def\PYG@tc##1{\textcolor[rgb]{0.40,0.40,0.40}{##1}}}
\expandafter\def\csname PYG@tok@gp\endcsname{\let\PYG@bf=\textbf\def\PYG@tc##1{\textcolor[rgb]{0.00,0.00,0.50}{##1}}}
\expandafter\def\csname PYG@tok@sh\endcsname{\def\PYG@tc##1{\textcolor[rgb]{0.73,0.13,0.13}{##1}}}
\expandafter\def\csname PYG@tok@ow\endcsname{\let\PYG@bf=\textbf\def\PYG@tc##1{\textcolor[rgb]{0.67,0.13,1.00}{##1}}}
\expandafter\def\csname PYG@tok@sx\endcsname{\def\PYG@tc##1{\textcolor[rgb]{0.00,0.50,0.00}{##1}}}
\expandafter\def\csname PYG@tok@bp\endcsname{\def\PYG@tc##1{\textcolor[rgb]{0.00,0.50,0.00}{##1}}}
\expandafter\def\csname PYG@tok@c1\endcsname{\let\PYG@it=\textit\def\PYG@tc##1{\textcolor[rgb]{0.25,0.50,0.50}{##1}}}
\expandafter\def\csname PYG@tok@fm\endcsname{\def\PYG@tc##1{\textcolor[rgb]{0.00,0.00,1.00}{##1}}}
\expandafter\def\csname PYG@tok@o\endcsname{\def\PYG@tc##1{\textcolor[rgb]{0.40,0.40,0.40}{##1}}}
\expandafter\def\csname PYG@tok@kc\endcsname{\let\PYG@bf=\textbf\def\PYG@tc##1{\textcolor[rgb]{0.00,0.50,0.00}{##1}}}
\expandafter\def\csname PYG@tok@c\endcsname{\let\PYG@it=\textit\def\PYG@tc##1{\textcolor[rgb]{0.25,0.50,0.50}{##1}}}
\expandafter\def\csname PYG@tok@mf\endcsname{\def\PYG@tc##1{\textcolor[rgb]{0.40,0.40,0.40}{##1}}}
\expandafter\def\csname PYG@tok@err\endcsname{\def\PYG@bc##1{\setlength{\fboxsep}{0pt}\fcolorbox[rgb]{1.00,0.00,0.00}{1,1,1}{\strut ##1}}}
\expandafter\def\csname PYG@tok@mb\endcsname{\def\PYG@tc##1{\textcolor[rgb]{0.40,0.40,0.40}{##1}}}
\expandafter\def\csname PYG@tok@ss\endcsname{\def\PYG@tc##1{\textcolor[rgb]{0.10,0.09,0.49}{##1}}}
\expandafter\def\csname PYG@tok@sr\endcsname{\def\PYG@tc##1{\textcolor[rgb]{0.73,0.40,0.53}{##1}}}
\expandafter\def\csname PYG@tok@mo\endcsname{\def\PYG@tc##1{\textcolor[rgb]{0.40,0.40,0.40}{##1}}}
\expandafter\def\csname PYG@tok@kd\endcsname{\let\PYG@bf=\textbf\def\PYG@tc##1{\textcolor[rgb]{0.00,0.50,0.00}{##1}}}
\expandafter\def\csname PYG@tok@mi\endcsname{\def\PYG@tc##1{\textcolor[rgb]{0.40,0.40,0.40}{##1}}}
\expandafter\def\csname PYG@tok@kn\endcsname{\let\PYG@bf=\textbf\def\PYG@tc##1{\textcolor[rgb]{0.00,0.50,0.00}{##1}}}
\expandafter\def\csname PYG@tok@cpf\endcsname{\let\PYG@it=\textit\def\PYG@tc##1{\textcolor[rgb]{0.25,0.50,0.50}{##1}}}
\expandafter\def\csname PYG@tok@kr\endcsname{\let\PYG@bf=\textbf\def\PYG@tc##1{\textcolor[rgb]{0.00,0.50,0.00}{##1}}}
\expandafter\def\csname PYG@tok@s\endcsname{\def\PYG@tc##1{\textcolor[rgb]{0.73,0.13,0.13}{##1}}}
\expandafter\def\csname PYG@tok@kp\endcsname{\def\PYG@tc##1{\textcolor[rgb]{0.00,0.50,0.00}{##1}}}
\expandafter\def\csname PYG@tok@w\endcsname{\def\PYG@tc##1{\textcolor[rgb]{0.73,0.73,0.73}{##1}}}
\expandafter\def\csname PYG@tok@kt\endcsname{\def\PYG@tc##1{\textcolor[rgb]{0.69,0.00,0.25}{##1}}}
\expandafter\def\csname PYG@tok@sc\endcsname{\def\PYG@tc##1{\textcolor[rgb]{0.73,0.13,0.13}{##1}}}
\expandafter\def\csname PYG@tok@sb\endcsname{\def\PYG@tc##1{\textcolor[rgb]{0.73,0.13,0.13}{##1}}}
\expandafter\def\csname PYG@tok@sa\endcsname{\def\PYG@tc##1{\textcolor[rgb]{0.73,0.13,0.13}{##1}}}
\expandafter\def\csname PYG@tok@k\endcsname{\let\PYG@bf=\textbf\def\PYG@tc##1{\textcolor[rgb]{0.00,0.50,0.00}{##1}}}
\expandafter\def\csname PYG@tok@se\endcsname{\let\PYG@bf=\textbf\def\PYG@tc##1{\textcolor[rgb]{0.73,0.40,0.13}{##1}}}
\expandafter\def\csname PYG@tok@sd\endcsname{\let\PYG@it=\textit\def\PYG@tc##1{\textcolor[rgb]{0.73,0.13,0.13}{##1}}}

\def\PYGZus{\char`\_}
\def\PYGZob{\char`\{}
\def\PYGZcb{\char`\}}

\def\PYGZam{\char`\&}
\def\PYGZlt{\char`\<}
\def\PYGZgt{\char`\>}

\def\PYGZpc{\char`\%}

\def\PYGZhy{\char`\-}

\def\PYGZdq{\char`\"}

\makeatother

\makeatletter
\def\PYGdefault@reset{\let\PYGdefault@it=\relax \let\PYGdefault@bf=\relax%
    \let\PYGdefault@ul=\relax \let\PYGdefault@tc=\relax%
    \let\PYGdefault@bc=\relax \let\PYGdefault@ff=\relax}
\def\PYGdefault@tok#1{\csname PYGdefault@tok@#1\endcsname}
\def\PYGdefault@toks#1+{\ifx\relax#1\empty\else%
    \PYGdefault@tok{#1}\expandafter\PYGdefault@toks\fi}
\def\PYGdefault@do#1{\PYGdefault@bc{\PYGdefault@tc{\PYGdefault@ul{%
    \PYGdefault@it{\PYGdefault@bf{\PYGdefault@ff{#1}}}}}}}
\def\PYGdefault#1#2{\PYGdefault@reset\PYGdefault@toks#1+\relax+\PYGdefault@do{#2}}

\expandafter\def\csname PYGdefault@tok@gd\endcsname{\def\PYGdefault@tc##1{\textcolor[rgb]{0.63,0.00,0.00}{##1}}}
\expandafter\def\csname PYGdefault@tok@gu\endcsname{\let\PYGdefault@bf=\textbf\def\PYGdefault@tc##1{\textcolor[rgb]{0.50,0.00,0.50}{##1}}}
\expandafter\def\csname PYGdefault@tok@gt\endcsname{\def\PYGdefault@tc##1{\textcolor[rgb]{0.00,0.27,0.87}{##1}}}
\expandafter\def\csname PYGdefault@tok@gs\endcsname{\let\PYGdefault@bf=\textbf}
\expandafter\def\csname PYGdefault@tok@gr\endcsname{\def\PYGdefault@tc##1{\textcolor[rgb]{1.00,0.00,0.00}{##1}}}
\expandafter\def\csname PYGdefault@tok@cm\endcsname{\let\PYGdefault@it=\textit\def\PYGdefault@tc##1{\textcolor[rgb]{0.25,0.50,0.50}{##1}}}
\expandafter\def\csname PYGdefault@tok@vg\endcsname{\def\PYGdefault@tc##1{\textcolor[rgb]{0.10,0.09,0.49}{##1}}}
\expandafter\def\csname PYGdefault@tok@vi\endcsname{\def\PYGdefault@tc##1{\textcolor[rgb]{0.10,0.09,0.49}{##1}}}
\expandafter\def\csname PYGdefault@tok@vm\endcsname{\def\PYGdefault@tc##1{\textcolor[rgb]{0.10,0.09,0.49}{##1}}}
\expandafter\def\csname PYGdefault@tok@mh\endcsname{\def\PYGdefault@tc##1{\textcolor[rgb]{0.40,0.40,0.40}{##1}}}
\expandafter\def\csname PYGdefault@tok@cs\endcsname{\let\PYGdefault@it=\textit\def\PYGdefault@tc##1{\textcolor[rgb]{0.25,0.50,0.50}{##1}}}
\expandafter\def\csname PYGdefault@tok@ge\endcsname{\let\PYGdefault@it=\textit}
\expandafter\def\csname PYGdefault@tok@vc\endcsname{\def\PYGdefault@tc##1{\textcolor[rgb]{0.10,0.09,0.49}{##1}}}
\expandafter\def\csname PYGdefault@tok@il\endcsname{\def\PYGdefault@tc##1{\textcolor[rgb]{0.40,0.40,0.40}{##1}}}
\expandafter\def\csname PYGdefault@tok@go\endcsname{\def\PYGdefault@tc##1{\textcolor[rgb]{0.53,0.53,0.53}{##1}}}
\expandafter\def\csname PYGdefault@tok@cp\endcsname{\def\PYGdefault@tc##1{\textcolor[rgb]{0.74,0.48,0.00}{##1}}}
\expandafter\def\csname PYGdefault@tok@gi\endcsname{\def\PYGdefault@tc##1{\textcolor[rgb]{0.00,0.63,0.00}{##1}}}
\expandafter\def\csname PYGdefault@tok@gh\endcsname{\let\PYGdefault@bf=\textbf\def\PYGdefault@tc##1{\textcolor[rgb]{0.00,0.00,0.50}{##1}}}
\expandafter\def\csname PYGdefault@tok@ni\endcsname{\let\PYGdefault@bf=\textbf\def\PYGdefault@tc##1{\textcolor[rgb]{0.60,0.60,0.60}{##1}}}
\expandafter\def\csname PYGdefault@tok@nl\endcsname{\def\PYGdefault@tc##1{\textcolor[rgb]{0.63,0.63,0.00}{##1}}}
\expandafter\def\csname PYGdefault@tok@nn\endcsname{\let\PYGdefault@bf=\textbf\def\PYGdefault@tc##1{\textcolor[rgb]{0.00,0.00,1.00}{##1}}}
\expandafter\def\csname PYGdefault@tok@no\endcsname{\def\PYGdefault@tc##1{\textcolor[rgb]{0.53,0.00,0.00}{##1}}}
\expandafter\def\csname PYGdefault@tok@na\endcsname{\def\PYGdefault@tc##1{\textcolor[rgb]{0.49,0.56,0.16}{##1}}}
\expandafter\def\csname PYGdefault@tok@nb\endcsname{\def\PYGdefault@tc##1{\textcolor[rgb]{0.00,0.50,0.00}{##1}}}
\expandafter\def\csname PYGdefault@tok@nc\endcsname{\let\PYGdefault@bf=\textbf\def\PYGdefault@tc##1{\textcolor[rgb]{0.00,0.00,1.00}{##1}}}
\expandafter\def\csname PYGdefault@tok@nd\endcsname{\def\PYGdefault@tc##1{\textcolor[rgb]{0.67,0.13,1.00}{##1}}}
\expandafter\def\csname PYGdefault@tok@ne\endcsname{\let\PYGdefault@bf=\textbf\def\PYGdefault@tc##1{\textcolor[rgb]{0.82,0.25,0.23}{##1}}}
\expandafter\def\csname PYGdefault@tok@nf\endcsname{\def\PYGdefault@tc##1{\textcolor[rgb]{0.00,0.00,1.00}{##1}}}
\expandafter\def\csname PYGdefault@tok@si\endcsname{\let\PYGdefault@bf=\textbf\def\PYGdefault@tc##1{\textcolor[rgb]{0.73,0.40,0.53}{##1}}}
\expandafter\def\csname PYGdefault@tok@s2\endcsname{\def\PYGdefault@tc##1{\textcolor[rgb]{0.73,0.13,0.13}{##1}}}
\expandafter\def\csname PYGdefault@tok@nt\endcsname{\let\PYGdefault@bf=\textbf\def\PYGdefault@tc##1{\textcolor[rgb]{0.00,0.50,0.00}{##1}}}
\expandafter\def\csname PYGdefault@tok@nv\endcsname{\def\PYGdefault@tc##1{\textcolor[rgb]{0.10,0.09,0.49}{##1}}}
\expandafter\def\csname PYGdefault@tok@s1\endcsname{\def\PYGdefault@tc##1{\textcolor[rgb]{0.73,0.13,0.13}{##1}}}
\expandafter\def\csname PYGdefault@tok@dl\endcsname{\def\PYGdefault@tc##1{\textcolor[rgb]{0.73,0.13,0.13}{##1}}}
\expandafter\def\csname PYGdefault@tok@ch\endcsname{\let\PYGdefault@it=\textit\def\PYGdefault@tc##1{\textcolor[rgb]{0.25,0.50,0.50}{##1}}}
\expandafter\def\csname PYGdefault@tok@m\endcsname{\def\PYGdefault@tc##1{\textcolor[rgb]{0.40,0.40,0.40}{##1}}}
\expandafter\def\csname PYGdefault@tok@gp\endcsname{\let\PYGdefault@bf=\textbf\def\PYGdefault@tc##1{\textcolor[rgb]{0.00,0.00,0.50}{##1}}}
\expandafter\def\csname PYGdefault@tok@sh\endcsname{\def\PYGdefault@tc##1{\textcolor[rgb]{0.73,0.13,0.13}{##1}}}
\expandafter\def\csname PYGdefault@tok@ow\endcsname{\let\PYGdefault@bf=\textbf\def\PYGdefault@tc##1{\textcolor[rgb]{0.67,0.13,1.00}{##1}}}
\expandafter\def\csname PYGdefault@tok@sx\endcsname{\def\PYGdefault@tc##1{\textcolor[rgb]{0.00,0.50,0.00}{##1}}}
\expandafter\def\csname PYGdefault@tok@bp\endcsname{\def\PYGdefault@tc##1{\textcolor[rgb]{0.00,0.50,0.00}{##1}}}
\expandafter\def\csname PYGdefault@tok@c1\endcsname{\let\PYGdefault@it=\textit\def\PYGdefault@tc##1{\textcolor[rgb]{0.25,0.50,0.50}{##1}}}
\expandafter\def\csname PYGdefault@tok@fm\endcsname{\def\PYGdefault@tc##1{\textcolor[rgb]{0.00,0.00,1.00}{##1}}}
\expandafter\def\csname PYGdefault@tok@o\endcsname{\def\PYGdefault@tc##1{\textcolor[rgb]{0.40,0.40,0.40}{##1}}}
\expandafter\def\csname PYGdefault@tok@kc\endcsname{\let\PYGdefault@bf=\textbf\def\PYGdefault@tc##1{\textcolor[rgb]{0.00,0.50,0.00}{##1}}}
\expandafter\def\csname PYGdefault@tok@c\endcsname{\let\PYGdefault@it=\textit\def\PYGdefault@tc##1{\textcolor[rgb]{0.25,0.50,0.50}{##1}}}
\expandafter\def\csname PYGdefault@tok@mf\endcsname{\def\PYGdefault@tc##1{\textcolor[rgb]{0.40,0.40,0.40}{##1}}}
\expandafter\def\csname PYGdefault@tok@err\endcsname{\def\PYGdefault@bc##1{\setlength{\fboxsep}{0pt}\fcolorbox[rgb]{1.00,0.00,0.00}{1,1,1}{\strut ##1}}}
\expandafter\def\csname PYGdefault@tok@mb\endcsname{\def\PYGdefault@tc##1{\textcolor[rgb]{0.40,0.40,0.40}{##1}}}
\expandafter\def\csname PYGdefault@tok@ss\endcsname{\def\PYGdefault@tc##1{\textcolor[rgb]{0.10,0.09,0.49}{##1}}}
\expandafter\def\csname PYGdefault@tok@sr\endcsname{\def\PYGdefault@tc##1{\textcolor[rgb]{0.73,0.40,0.53}{##1}}}
\expandafter\def\csname PYGdefault@tok@mo\endcsname{\def\PYGdefault@tc##1{\textcolor[rgb]{0.40,0.40,0.40}{##1}}}
\expandafter\def\csname PYGdefault@tok@kd\endcsname{\let\PYGdefault@bf=\textbf\def\PYGdefault@tc##1{\textcolor[rgb]{0.00,0.50,0.00}{##1}}}
\expandafter\def\csname PYGdefault@tok@mi\endcsname{\def\PYGdefault@tc##1{\textcolor[rgb]{0.40,0.40,0.40}{##1}}}
\expandafter\def\csname PYGdefault@tok@kn\endcsname{\let\PYGdefault@bf=\textbf\def\PYGdefault@tc##1{\textcolor[rgb]{0.00,0.50,0.00}{##1}}}
\expandafter\def\csname PYGdefault@tok@cpf\endcsname{\let\PYGdefault@it=\textit\def\PYGdefault@tc##1{\textcolor[rgb]{0.25,0.50,0.50}{##1}}}
\expandafter\def\csname PYGdefault@tok@kr\endcsname{\let\PYGdefault@bf=\textbf\def\PYGdefault@tc##1{\textcolor[rgb]{0.00,0.50,0.00}{##1}}}
\expandafter\def\csname PYGdefault@tok@s\endcsname{\def\PYGdefault@tc##1{\textcolor[rgb]{0.73,0.13,0.13}{##1}}}
\expandafter\def\csname PYGdefault@tok@kp\endcsname{\def\PYGdefault@tc##1{\textcolor[rgb]{0.00,0.50,0.00}{##1}}}
\expandafter\def\csname PYGdefault@tok@w\endcsname{\def\PYGdefault@tc##1{\textcolor[rgb]{0.73,0.73,0.73}{##1}}}
\expandafter\def\csname PYGdefault@tok@kt\endcsname{\def\PYGdefault@tc##1{\textcolor[rgb]{0.69,0.00,0.25}{##1}}}
\expandafter\def\csname PYGdefault@tok@sc\endcsname{\def\PYGdefault@tc##1{\textcolor[rgb]{0.73,0.13,0.13}{##1}}}
\expandafter\def\csname PYGdefault@tok@sb\endcsname{\def\PYGdefault@tc##1{\textcolor[rgb]{0.73,0.13,0.13}{##1}}}
\expandafter\def\csname PYGdefault@tok@sa\endcsname{\def\PYGdefault@tc##1{\textcolor[rgb]{0.73,0.13,0.13}{##1}}}
\expandafter\def\csname PYGdefault@tok@k\endcsname{\let\PYGdefault@bf=\textbf\def\PYGdefault@tc##1{\textcolor[rgb]{0.00,0.50,0.00}{##1}}}
\expandafter\def\csname PYGdefault@tok@se\endcsname{\let\PYGdefault@bf=\textbf\def\PYGdefault@tc##1{\textcolor[rgb]{0.73,0.40,0.13}{##1}}}
\expandafter\def\csname PYGdefault@tok@sd\endcsname{\let\PYGdefault@it=\textit\def\PYGdefault@tc##1{\textcolor[rgb]{0.73,0.13,0.13}{##1}}}

\makeatother

\usepackage[final,nonatbib]{nips_2017}

\usepackage[utf8]{inputenc} % allow utf-8 input
\usepackage[T1]{fontenc}    % use 8-bit T1 fonts
\usepackage{hyperref}       % hyperlinks
\usepackage{url}            % simple URL typesetting
\usepackage{booktabs}       % professional-quality tables
\usepackage{amsfonts}       % blackboard math symbols
\usepackage{nicefrac}       % compact symbols for 1/2, etc.
\usepackage{microtype}      % microtypography

\usepackage{adjustbox}
\usepackage{array}

\newcolumntype{R}[2]{%
  >{\adjustbox{angle=#1,lap=\width-(#2)}\bgroup}%
  l%
  <{\egroup}%
}
\newcommand*\rot{\multicolumn{1}{R{30}{2em}}}

\title{A generic and fast C++ optimization framework}

\author{
  Ryan R. Curtin \\
  Symantec Corporation \\
  Atlanta, GA, USA 30318 \\
  \texttt{ryan@ratml.org} \\
  \And
  Shikhar Bhardwaj \\
  Delhi Technological University \\
  Delhi - 110042 \\
  \texttt{shikhar\char`_bt2k15@dtu.ac.in}
  \And
  Marcus Edel \\
  Free University of Berlin \\
  Arnimallee 7, 14195 Berlin \\
  \texttt{marcus.edel@fu-berlin.de} \\
  \And
  Yannis Mentekidis \\
  mentekid@gmail.com
}

\begin{document}
% \nipsfinalcopy is no longer used

\maketitle

\vspace*{-0.5cm}
\begin{abstract}
\vspace*{-0.3em}
The development of the {\bf mlpack} C++ machine learning library
(\url{http://www.mlpack.org/}) has required the design and implementation of a
flexible, robust optimization system that is able to solve the types of
arbitrary optimization problems that may arise all throughout machine learning
problems.  In this paper, we present the generic optimization framework that we
have designed for mlpack.  A key priority in the design was ease of
implementation of both new optimizers and new objective functions to be
optimized; therefore, implementation of a new optimizer requires only one
method and implementation of a new objective function requires at most four
functions.  This leads to simple and intuitive code, which, for fast prototyping
and experimentation, is of paramount importance.  When compared to
optimization frameworks of other libraries, we find that mlpack's supports more
types of objective functions, is able to make optimizations that other
frameworks do not, and seamlessly supports user-defined objective functions
{\it and} optimizers.
%\begin{itemize}
%  \item It's easy to implement new optimizers.
%  \item It's easy to implement new functions to be optimized.
%  \item This system is more comprehensive than other techniques that may focus
%only on neural networks or single problems.
%  \item This system is also more specific and offers more room for optimization
%than super general systems like scipy.optimize().
%\end{itemize}
%
%This abstract will need to be rewritten.
\end{abstract}

\vspace*{-0.7cm}
\section{Introduction}
\vspace*{-0.3em}

Machine learning is a field that is inextricably intertwined with the field of
optimization.  Countless machine learning techniques depend on the optimization
of a given objective function; for instance, classifiers such as logistic
regression \cite{cox1958regression}, metric learning methods like NCA
\cite{goldberger2005neighbourhood}, manifold learning algorithms like MVU
\cite{weinberger2006introduction}, and the extremely popular field of deep
learning \cite{schmidhuber2015deep}.  Thanks to the attention focused on these
problems, it is increasingly important in the field to have fast, practical
optimizers.  This can explain the current focus on optimization: this year at
ICML (2017), every single session had at least one track devoted solely to
optimization techniques.

Therefore, the need is real to provide a robust, flexible framework in which
new optimizers can be easily developed.  Similarly, the need is also real for a
flexible framework that allows new objective functions to be easily implemented
and optimized with a variety of possible optimizers.

However, the current landscape of optimization frameworks for machine learning
is not particularly comprehensive.  A variety of tools such as
Caffe \cite{jia2014caffe},
% mxnet ??
TensorFlow \cite{abadi2016tensorflow},
and
Keras \cite{chollet2015}
have optimization frameworks, but they are limited to SGD-type optimizers and
are only able to optimize deep neural networks or related structures.  Thus
expressing arbitrary machine learning objective functions can be difficult or in
some cases not possible.  Other libraries, like
% Shogun \cite{shogun}
%and
scikit-learn \cite{pedregosa2011scikit},
do have optimizers, but generally not in a coherent framework and
often the implementations may be specific to an individual machine learning
algorithm.  At a higher level, many programming languages may have generic
optimizers, like SciPy \cite{jones2014scipy} and MATLAB \cite{mathworks2017OTB}, but typically these
optimizers are not suitable for large-scale machine learning tasks where, e.g.,
calculating the full gradient of all of the data may not be feasible.  Table
\ref{tab:features} provides a rough overview.

Given this situation, we have developed a flexible optimization infrastructure
in the {\bf mlpack} C++ machine learning library \cite{mlpack2013}.  This
infrastructure makes it easy to combine nearly any type of optimizer with nearly
any type of objective function, and has allowed us to minimize the effort
necessary to both implement new optimizers and to implement new machine learning
algorithms that depend on optimization.  Since the framework is implemented in
C++ and uses template metaprogramming, we are able to preserve clean syntax
while simultaneously allowing compile-time speedups that can result in a
generic optimizer that is as fast as an optimizer written for one specific
objective function.

In this short paper, we describe mlpack's optimization infrastructure in detail.
First, we describe the types of optimization problems we would like to solve,
which then allows us to build a generic and intuitive API for objective
functions to implement and optimizers to depend upon.  We show a very simple
example objective function and optimizer, and a system that we have built to
detect when an objective function that cannot be optimized by a certain
optimizer is used.  Lastly, we show the optimizers we have implemented in this
framework, and example usage of mlpack's optimizers.

%In recent times, machine learning conferences have focused more and more on
%optimization techniques.  For instance, this year at ICML (ICML 2017), every
%single session had at least one track devoted solely to optimization
%techniques.
%<TODO: add some more filler here.  Show that optimization is an important part
%of machine learning, in whatever ways.>

%\begin{itemize}
%  \item This sequence of events (or situation) motivates the development of a
%generic, flexible, and fast optimization framework.
%  \item We have implemented exactly this in {\bf mlpack}, and it's good.
%\end{itemize}

%{\bf <TODO: elaborate on above, add a few more sentences.>}

%{\bf <TODO: mention related work, or really that we haven't found other
%libraries that have a comprehensive infrastructure like the one we have built,
%and that what we have built is more flexible and generic}

% Just to get a feeling of the look; Also, we could also just use \OK instead of \CURCLE or \LEFTcircle
\begin{table}
\centering
    \begin{tabular}{@{} cl*{7}c @{}}
%  \begin{tabular}{ccccccc}
        & & \multicolumn{7}{c}{} \\[2ex]
            % If there is any coherent framework at all, this is true.
        & & \rot{has framework}
            % If there is any support for constrained optimization, this is
            % true.
          & \rot{constraints}
            % If the optimization framework can do mini-batch, this is true.
          & \rot{batches}
            % If I can implement any arbitrary function to be optimized, this is
            % true.
          & \rot{arbitrary functions}
            % If I can implement any new optimization technique to use, this is
            % true.
          & \rot{arbitrary optimizers}
            % If the framework could take advantage of when the gradient is
            % sparse, this is true.
          & \rot{sparse gradients}
            % If the framework can handle categorical/discrete variables for
            % optimization, this is true.
          & \rot{categorical} \\
        \cmidrule{2-9}
  \hline
        % It might be reasonable to say mlpack categorical support is only
        % partial, but I am not sure exactly where we draw the line.
        & mlpack \cite{mlpack2013}            & \CIRCLE & \CIRCLE & \CIRCLE & \CIRCLE & \CIRCLE & \CIRCLE & \CIRCLE \\
        % The Shogun toolbox has a fairly nice framework, but it doesn't support
        % sparse gradients or categorical features.  It also does not appear to
        % support constraints.
        & Shogun \cite{sonnenburg2010shogun}             & \CIRCLE & - & \CIRCLE & \CIRCLE & \CIRCLE & - & -
\\
        % VW doesn't appear to have any framework whatsoever and the code is
        % awful, but it does support batches and categorical features.
        & Vowpal Wabbit \cite{Langford2007VW}      & - & - & \CIRCLE  & - & - & - &
\CIRCLE \\
        % TensorFlow has a few optimizers, but they are all SGD-related.  You
        % can write most objectives easily (but some very hard), and categorical
        % support might be possible but would not be easy.
        & TensorFlow \cite{abadi2016tensorflow}        & \CIRCLE & -  & \CIRCLE  & \LEFTcircle & - &
\LEFTcircle & -  \\
        % Caffe has a nice framework, but it's only for SGD-related optimizers.
        % I think I could write a new one, but it is not the easiest thing in
        % the world.
        & Caffe \cite{jia2014caffe}           & \CIRCLE & -  & \CIRCLE & \LEFTcircle & \LEFTcircle
& - & - \\
        % Keras is restricted to neural networks and SGD-like optimizers.  I
        % don't know that it is possible to easily write a new optimizer.
        & Keras \cite{chollet2015}            & \CIRCLE & -  & \CIRCLE & \LEFTcircle & \LEFTcircle
& - & - \\
        % sklearn has a few optimizer frameworks, but they are all in different
        % places and have somewhat different support.
        & scikit-learn \cite{pedregosa2011scikit}       & \LEFTcircle & - & \LEFTcircle  & \LEFTcircle & -
& - & - \\
        % scipy has some nice optimizer framework but it does not support
        % batches or some of the more complex functionality.  And you can't
        % write your own.
        & SciPy \cite{jones2014scipy}             & \CIRCLE & \CIRCLE  & -  & \CIRCLE & - & - & - \\
        % MATLAB is very similar to scipy.
        & MATLAB \cite{mathworks2017OTB}            & \CIRCLE & \CIRCLE & - & \CIRCLE & - & - & - \\
        \cmidrule[1pt]{2-9}
    \end{tabular}
%   \begin{tablenotes}\footnotesize
\caption{
Feature comparison: \CIRCLE = provides feature,
\LEFTcircle = partially provides feature, - = does not provide feature.
{\it has framework} the library has some kind of generic
optimization framework; {\it constraints} and {\it batches} indicate support for
constrained problems and batches; {\it arbitrary functions} means arbitrary
objective functions are easily implemented; {\it arbitrary optimizers} means
arbitrary optimizers are easily implemented; {\it sparse gradient} indicates
that the framework can natively take advantage of sparse gradients; and
{\it categorical} refers to if categorical feature support exists.
}
\label{tab:features}
\vspace*{-1.8em}
\end{table}

\vspace*{-0.3em}
\section{The mlpack machine learning library}
\vspace*{-0.2em}

mlpack is a C++ machine learning library with an emphasis on speed, flexibility,
and ease of use \cite{mlpack2013}; it has been continuously developed since
2007.  mlpack uses the Armadillo library \cite{pasc2017} for
linear algebra and matrix primitives.  Both mlpack and Armadillo exploit C++
language features such as policy-based design and template metaprogramming to
provide compile-time optimizations that result in fast code \cite{pasc2017},
while retaining a simple and easy-to-use interface \cite{icopust2017}.

Many of the speedups in mlpack depend on the technique of {\it policy-based
design} \cite{Alexandrescu2001}.  With this design paradigm, mlpack
provides a number of classes with modular functionality.  As a simple example,
mlpack's mean shift clustering implementation (as of version 2.2.5) takes three
template parameters:

\vspace*{-0.4em}
{\footnotesize
\begin{Verbatim}[commandchars=\\\{\}]
  \PYG{k}{template}\PYG{o}{\PYGZlt{}}\PYG{k+kt}{bool} \PYG{n}{UseKernel}\PYG{p}{,} \PYG{k}{typename} \PYG{n}{KernelType}\PYG{p}{,} \PYG{k}{typename} \PYG{n}{MatType}\PYG{o}{\PYGZgt{}}
  \PYG{k}{class} \PYG{n+nc}{MeanShift}\PYG{p}{;}
\end{Verbatim}
}
\vspace*{-0.4em}

Thus a user wishing to perform kernelized mean shift clustering with the
Gaussian kernel might simply use the class
{\footnotesize {\tt MeanShift<true, GaussianKernel, arma::mat>}}
%\begin{Verbatim}[commandchars=\\\{\}]
%\PYG{n}{MeanShift}\PYG{o}{\PYGZlt{}}\PYG{n+nb}{true}\PYG{p}{,} \PYG{n}{GaussianKernel}\PYG{p}{,} \PYG{n}{arma}\PYG{o}{::}\PYG{n}{mat}\PYG{o}{\PYGZgt{}}
%\end{Verbatim}
%}
\noindent where {\tt arma::mat} is Armadillo's dense matrix type and {\tt
GaussianKernel} provides one simple {\tt Evaluate()} method.

Since the {\tt KernelType} argument is a template argument, any class can be
used---it does not need to be part of mlpack.  Therefore, a user can simply
implement their own {\tt KernelType} class with the single {\tt Evaluate()}
method and use that as a template argument to the {\tt MeanShift} class, and
the compiler will generate a specialized implementation using the custom
kernel.  Similar support could be accomplished by, e.g., function pointers (and
this is often done in other languages); however, templates allow us to avoid
the runtime cost of dereferencing the function pointer by generating code that
directly calls the correct method. This cost can be non-negligible, especially
in situations where the method is called many times, such as the {\tt
Evaluate()} method of an optimization problem.

mlpack uses policy-based design throughout the library, so that any
functionality can be easily extended or modified by the user without needing to
dig into the internals of the code.  Our optimization framework is built around
this paradigm, allowing for fast prototyping of either new optimizers or new
objective functions.

\vspace*{-0.3em}
\section{Requirements for optimizers and functions}
\vspace*{-0.2em}

In general, we want to be able to consider any solver of the problem

\vspace*{-0.5em}
\begin{equation}
\operatorname{argmin}_{x} f(x)
\end{equation}
\vspace*{-1.3em}

\noindent for any function $f(x)$ that takes some vector input $x$.  But it
is impossible to implement something both so generic and fast---for instance,
gradient-based approaches converge far more quickly than gradient-free
approaches (in general), so we must design an abstraction that is able to
simultaneously generalize to many problem types, as well as take advantage of
accelerations and optimizations.

Let us describe the class of functions to be optimized with some non-exclusive
properties:

\vspace*{-0.4em}
\begin{itemize} \itemsep -1pt
  \item {\bf arbitrary}: no assumptions can be made on $f(x)$
  \item {\bf differentiable}: $f(x)$ has a computable gradient $f'(x)$
  \item {\bf separable}: $f(x)$ is a sum of individual components: $f(x) =
\sum_{i} f_i(x)$
  \item {\bf categorical}: $x$ contains elements that can only take discrete
values
  %\item {\bf numeric}: all elements of $x$ take values in $\mathcal{R}$
  \item {\bf sparse}: the gradient $f'(x)$ or $f_i(x)$ (for a separable
function) is sparse
  \item {\bf partially differentiable}: the gradient $f'_j(x)$ is computable for
individual elements $x_j$ of $x$
  \item {\bf bounded}: $x$ is limited in the values that it can take
\end{itemize}
\vspace*{-0.4em}

Needless to say, it is impossible to create a system where every optimizer can
work with every possible type of function: a gradient-based optimizer cannot
reasonably be expected to work with an arbitrary function $f(x)$ where the
gradient is not computable or available.

Instead, the best we can hope to achieve is to maximize the flexibility
available, so that a user can easily implement a function $f(x)$ and have it
work with as many optimizers as possible.  For this, C++ policy-based design
aids us greatly: when implementing a function to be optimized, a user can
implement only a few methods and we can use C++ template metaprogramming to
check that the given functions match the requirements of the optimizer that is
being used.  When implementing an optimizer, a user can assume that the given
function to be optimized meets the required assumptions of the optimizers, and
encode those requirements.
Since we are using templates and C++, the resulting
code generated by the compiler can be identical to what a developer would write
if they were writing an optimizer specifically for the function $f(x)$---this
can give significant efficiency gains.

In some cases, some subclasses of functions can still be optimized with more
general optimizers.  For example, separable functions can still be optimized
with an optimizer that does not specifically support them.  Similarly, a sparse
differentiable function may be optimized with any optimizer that supports
differentiable functions; however, the sparseness might not be taken advantage
of.

Now, with an understanding of the types of functions that we wish to support and
their individual characteristics, we can develop an API that a given function
can implement the relevant parts of.

\vspace*{-0.3em}
\section{{\tt FunctionType} API}
\vspace*{-0.2em}

In order to facilitate consistent implementations, we have defined a {\tt
FunctionType} API that describes all the methods that an objective function
may implement.  mlpack offers a few variations of this API, each designed to
cover some of the function characteristics of the previous section.
%At the time of writing, we offer a few implementations of this API, each
%designed to cover a special case.
%\begin{itemize}
%  \item{\tt FunctionType}: Requires the implementation of {\tt Evaluate()} and
%  {\tt Gradient()}. This is a generic function that is used by the
%  {\tt GradientDescent} optimizer.
%  \item{\tt DecomposableFunctionType}: In addition to {\tt FunctionType}'s
%  requirements, requires the implementation of {NumFunctions()}, which returns
%  the number of parts the optimization problem can be decomposed into. The other
%  two functions ({\tt Gradient(), Evaluate()}) accept an index as argument and
%  only evaluate one of the decomposable parts. This is used by our {\tt AdaGrad}
%  implementations, some {\tt SGD} implementations, {\tt RMSProp} and
%  {\tt SMORMS3}.
%  \item{\tt SparseFunctionType}: Similar to {\tt DecomposableFunctionType},
%  however this function type returns a sparse gradient, which makes it ideal for
%  sparse, parallelizable optimization problems. This function type is used by
%  {\bf mlpack}'s parallel implementation of SGD, {\tt ParallelSGD}.
% We'll kind of sweep constrained optimization under the rug a little bit, there
% are a lot of extra complexities we don't have space for.
%  \item{\tt LagrangianFunctionType}: This function type fits constrained
%  optimization problems. It requires the implementation of {\tt NumConstraints},
%  {\tt EvaluateConstraint()} and {\tt GradientConstraint}, which are required by
%  optimizers like {\tt AugLagrangian} and {\tt LRSDP}.
%\end{itemize}
Any {\tt FunctionType} to be optimized requires the implementation of an {\tt
Evaluate()} method.  The interface used for that can be one of the following
two methods:

\vspace*{-0.3em}
{\footnotesize
\begin{Verbatim}[commandchars=\\\{\}]
\PYG{c+c1}{// For non\PYGZhy{}separable objectives.}
\PYG{k+kt}{double} \PYG{n+nf}{Evaluate}\PYG{p}{(}\PYG{k}{const} \PYG{n}{arma}\PYG{o}{::}\PYG{n}{mat}\PYG{o}{\PYGZam{}} \PYG{n}{parameters}\PYG{p}{);}
\PYG{c+c1}{// For separable objectives.}
\PYG{k+kt}{double} \PYG{n+nf}{Evaluate}\PYG{p}{(}\PYG{k}{const} \PYG{n}{arma}\PYG{o}{::}\PYG{n}{mat}\PYG{o}{\PYGZam{}} \PYG{n}{parameters}\PYG{p}{,}
                \PYG{k}{const} \PYG{k+kt}{size\PYGZus{}t} \PYG{n}{start}\PYG{p}{,}
                \PYG{k}{const} \PYG{k+kt}{size\PYGZus{}t} \PYG{n}{batchSize}\PYG{p}{);}
\end{Verbatim}
}
\vspace*{-0.3em}

Both of these methods should calculate the objective given the parameters
matrix {\tt parameters}; however, the second overload is for {\it separable}
functions, and should calculate the partial objective starting at the separable
function indexed by {\tt start} and calculate {\tt batchSize} partial objectives
and return the sum.  Functions implementing the first overload are used by
optimizers like the {\tt GradientDescent} optimizer; functions implementing the
second are used by SGD-like optimizers.

Note that, importantly, it is easy to calculate the objective for a
non-separable function with the second overload just by setting {\tt start} to
$0$ and {\tt batchSize} to $1$.  Therefore, it is easy to make an `adapter' that
can allow separable optimizers to work with non-separable functions (and vice
versa).

Next, any {\it differentiable} function must implement some {\tt Gradient()}
method.

\vspace*{-0.3em}
{\footnotesize
\begin{Verbatim}[commandchars=\\\{\}]
\PYG{c+c1}{// For non\PYGZhy{}separable differentiable sparse and non\PYGZhy{}sparse functions.}
\PYG{k}{template}\PYG{o}{\PYGZlt{}}\PYG{k}{typename} \PYG{n}{GradType}\PYG{o}{\PYGZgt{}}
\PYG{k+kt}{void} \PYG{n}{Gradient}\PYG{p}{(}\PYG{k}{const} \PYG{n}{arma}\PYG{o}{::}\PYG{n}{mat}\PYG{o}{\PYGZam{}} \PYG{n}{parameters}\PYG{p}{,} \PYG{n}{GradType}\PYG{o}{\PYGZam{}} \PYG{n}{gradient}\PYG{p}{);}
\PYG{c+c1}{// For separable differentiable sparse and non\PYGZhy{}sparse functions.}
\PYG{k}{template}\PYG{o}{\PYGZlt{}}\PYG{k}{typename} \PYG{n}{GradType}\PYG{o}{\PYGZgt{}}
\PYG{k+kt}{void} \PYG{n}{Gradient}\PYG{p}{(}\PYG{k}{const} \PYG{n}{arma}\PYG{o}{::}\PYG{n}{mat}\PYG{o}{\PYGZam{}} \PYG{n}{parameters}\PYG{p}{,}
              \PYG{k}{const} \PYG{k+kt}{size\PYGZus{}t} \PYG{n}{start}\PYG{p}{,}
              \PYG{n}{GradType}\PYG{o}{\PYGZam{}} \PYG{n}{gradient}\PYG{p}{,}
              \PYG{k}{const} \PYG{k+kt}{size\PYGZus{}t} \PYG{n}{batchSize}\PYG{p}{);}
\end{Verbatim}
}
\vspace*{-0.3em}

Both of these methods should calculate the gradient and place the results into
the matrix object {\tt gradient} that is passed as an argument.
Note that the method accepts a template
parameter {\tt GradType}, which may be {\tt arma::mat} (dense Armadillo matrix)
or {\tt arma::sp\_mat} (sparse Armadillo matrix).  This allows support for both
sparse-supporting and non-sparse-supporting optimizers.\footnote{One could write
a non-templated {\tt Gradient()} method for just {\tt arma::mat}, and it would
work fine for non-sparse optimizers.  But to us it seems just as easy to
templatize it.}

Next, if the objective function is {\it partially differentiable}, we can
implement the following method:

\vspace*{-0.3em}
{\footnotesize
\begin{Verbatim}[commandchars=\\\{\}]
\PYG{c+c1}{// For partially differentiable sparse and non\PYGZhy{}sparse functions.}
\PYG{k}{template}\PYG{o}{\PYGZlt{}}\PYG{k}{typename} \PYG{n}{GradType}\PYG{o}{\PYGZgt{}}
\PYG{k+kt}{void} \PYG{n}{PartialGradient}\PYG{p}{(}\PYG{k}{const} \PYG{n}{arma}\PYG{o}{::}\PYG{n}{mat}\PYG{o}{\PYGZam{}} \PYG{n}{parameters}\PYG{p}{,} \PYG{k}{const} \PYG{k+kt}{size\PYGZus{}t} \PYG{n}{j}\PYG{p}{,} \PYG{n}{GradType}\PYG{o}{\PYGZam{}} \PYG{n}{gradient}\PYG{p}{);}
\end{Verbatim}
}
\vspace*{-0.3em}

This should calculate the gradient of the parameters {\tt parameters} with
respect to the parameter {\tt j} and store the results (either sparse or dense)
in the {\tt gradient} matrix object.

In addition, {\it separable} and {\it partially differentiable} functions must
implement the {\tt NumFunctions()} and {\tt NumFeatures()} functions,
respectively:

\vspace*{-0.3em}
{\footnotesize
\begin{Verbatim}[commandchars=\\\{\}]
\PYG{c+c1}{// For separable functions: return the number of parts the optimization problem}
\PYG{c+c1}{// can be decomposed into.}
\PYG{k+kt}{size\PYGZus{}t} \PYG{n+nf}{NumFunctions}\PYG{p}{();}
\PYG{c+c1}{// For partially differentiable functions: return the number of partial derivatives.}
\PYG{k+kt}{size\PYGZus{}t} \PYG{n+nf}{NumFeatures}\PYG{p}{();}
\end{Verbatim}
}
\vspace*{-0.3em}

Finally, {\it separable} functions must implement the method
{\footnotesize {\tt void Shuffle()}},
%\begin{Verbatim}[commandchars=\\\{\}]
%\PYG{k+kt}{void} \PYG{n}{Shuffle}\PYG{p}{()}
%\end{Verbatim}
%},
which shuffles the ordering
of the functions (note that the optimizer is not required to call it).  This is
useful for data-based problems, where it may not be desirable to loop over the
separable objective functions---which usually correspond to individual data
points---in the same order.

These simple functions, however, do not specify how to handle {\it categorical}
or {\it bounded} functions.  In those cases, the {\tt FunctionType} should
accept, in its constructor, its bounds and which dimensions are categorical or
numeric.  Since constraint types can differ greatly, it is up to an individual
optimizer to define the format in which it should receive its constraints.  More
information can be found in the documentation of mlpack's {\tt LRSDP}, {\tt
AugLagrangian}, and {\tt FrankWolfe} optimizers.

%% The constraints are part of the optimizer themselves.
% EvaluateConstraint()
% NumConstraints()

%\begin{itemize}
%  \item {\tt double Evaluate(const arma::mat& parameters)} {\it (arbitrary
%functions)}: return the objective function given the parameters.
%  \item {\tt void Gradient(const arma::mat& parameters, arma::mat& gradient)}
%{\it (differentiable)}: compute the gradient and store it in {\tt gradient}
%given the parameters.
%  \item {\tt double Evaluate(const arma::mat& parameters, const size\_t start,
%const size\_t batchSize)} {\it (separable)}: compute a partial objective
%function, starting with separable function {\tt start} and computing {\tt batchSize}
%functions.
%  \item {\tt void Gradient(arma::mat& parameters, const size\_t start,
%arma::mat& gradient, const size\_t batchSize)} {\it (separable)}: compute a part
%of the gradient starting with separable function {\tt start} and computing {\tt
%batchSize} individual gradients, storing the result in {\tt gradient}.

\vspace*{-0.4em}
\section{Optimizer API}
\vspace*{-0.5em}

In addition to implementing functions, users can also add new optimizers easily
if they implement an optimizer with a simple API.  Fortunately, the requirements
for implementing optimizers are much simpler than for objective functions.  An
optimizer must implement only the method

\vspace*{-0.5em}
{\footnotesize
\begin{Verbatim}[commandchars=\\\{\}]
\PYG{k}{template}\PYG{o}{\PYGZlt{}}\PYG{k}{typename} \PYG{n}{FunctionType}\PYG{o}{\PYGZgt{}}
\PYG{k+kt}{double} \PYG{n}{Optimize}\PYG{p}{(}\PYG{n}{FunctionType}\PYG{o}{\PYGZam{}} \PYG{n}{function}\PYG{p}{,} \PYG{n}{arma}\PYG{o}{::}\PYG{n}{mat}\PYG{o}{\PYGZam{}} \PYG{n}{parameters}\PYG{p}{);}
\end{Verbatim}
}
\vspace*{-0.5em}

The {\tt Optimize()} method should check that the given {\tt FunctionType}
satisfies the assumptions the optimizer makes (see Section \ref{sec:static}) and
optimize the given function {\tt function}, storing the best set of parameters
in the matrix {\tt parameters} and returning the best objective value.

\vspace*{-0.6em}
\section{Example function and optimizer}
\vspace*{-0.7em}

This section details the usage of mlpack's optimization framework. In this
example, we would like to minimize a simple function, where each dimension has a
parabola with a distinct minimum. In this example, we show how to use mlpack’s
framework for this task, and minimize the function below.

% \vspace*{-0.3em}
%\begin{listing}[H]
\vspace*{-0.5em}
{\footnotesize
\begin{Verbatim}[commandchars=\\\{\}]
\PYG{k}{struct} \PYG{n}{ObjectiveFunction} \PYG{p}{\PYGZob{}}
  \PYG{n}{ObjectiveFunction}\PYG{p}{()} \PYG{p}{\PYGZob{}} \PYG{c+c1}{// A separable function consisting of four quadratics.}
    \PYG{n}{in} \PYG{o}{=} \PYG{n}{arma}\PYG{o}{::}\PYG{n}{vec}\PYG{p}{(}\PYG{l+s}{\PYGZdq{}20 12 15 100\PYGZdq{}}\PYG{p}{);} \PYG{n}{bi} \PYG{o}{=} \PYG{n}{arma}\PYG{o}{::}\PYG{n}{vec}\PYG{p}{(}\PYG{l+s}{\PYGZdq{}\PYGZhy{}4 \PYGZhy{}2 \PYGZhy{}3 \PYGZhy{}8\PYGZdq{}}\PYG{p}{);}
  \PYG{p}{\PYGZcb{}}

  \PYG{k+kt}{size\PYGZus{}t} \PYG{n}{NumFunctions}\PYG{p}{()} \PYG{p}{\PYGZob{}} \PYG{k}{return} \PYG{l+m+mi}{4}\PYG{p}{;} \PYG{p}{\PYGZcb{}}
  \PYG{k+kt}{void} \PYG{n}{Shuffle}\PYG{p}{()} \PYG{p}{\PYGZob{}} \PYG{n}{ord} \PYG{o}{=} \PYG{n}{arma}\PYG{o}{::}\PYG{n}{shuffle}\PYG{p}{(}\PYG{n}{arma}\PYG{o}{::}\PYG{n}{uvec}\PYG{p}{(}\PYG{l+s}{\PYGZdq{}0 1 2 3\PYGZdq{}}\PYG{p}{));} \PYG{p}{\PYGZcb{}}

  \PYG{k+kt}{double} \PYG{n}{Evaluate}\PYG{p}{(}\PYG{k}{const} \PYG{n}{arma}\PYG{o}{::}\PYG{n}{mat}\PYG{o}{\PYGZam{}} \PYG{n}{para}\PYG{p}{,} \PYG{k+kt}{size\PYGZus{}t} \PYG{n}{s}\PYG{p}{,} \PYG{k+kt}{size\PYGZus{}t} \PYG{n}{bs}\PYG{p}{)} \PYG{p}{\PYGZob{}}
    \PYG{k+kt}{double} \PYG{n}{cost} \PYG{o}{=} \PYG{l+m+mi}{0}\PYG{p}{;}
    \PYG{k}{for} \PYG{p}{(}\PYG{k+kt}{size\PYGZus{}t} \PYG{n}{i} \PYG{o}{=} \PYG{n}{s}\PYG{p}{;} \PYG{n}{i} \PYG{o}{\PYGZlt{}} \PYG{n}{s} \PYG{o}{+} \PYG{n}{bs}\PYG{p}{;} \PYG{n}{i}\PYG{o}{++}\PYG{p}{)}
      \PYG{n}{cost} \PYG{o}{+=} \PYG{n}{para}\PYG{p}{(}\PYG{n}{ord}\PYG{p}{[}\PYG{n}{i}\PYG{p}{])} \PYG{o}{*} \PYG{n}{para}\PYG{p}{(}\PYG{n}{ord}\PYG{p}{[}\PYG{n}{i}\PYG{p}{])} \PYG{o}{+} \PYG{n}{bi}\PYG{p}{(}\PYG{n}{ord}\PYG{p}{[}\PYG{n}{i}\PYG{p}{])} \PYG{o}{*} \PYG{n}{para}\PYG{p}{(}\PYG{n}{ord}\PYG{p}{[}\PYG{n}{i}\PYG{p}{])} \PYG{o}{+} \PYG{n}{in}\PYG{p}{(}\PYG{n}{ord}\PYG{p}{[}\PYG{n}{i}\PYG{p}{]);}
    \PYG{k}{return} \PYG{n}{cost}\PYG{p}{;}
  \PYG{p}{\PYGZcb{}}

  \PYG{k+kt}{void} \PYG{n}{Gradient}\PYG{p}{(}\PYG{k}{const} \PYG{n}{arma}\PYG{o}{::}\PYG{n}{mat}\PYG{o}{\PYGZam{}} \PYG{n}{para}\PYG{p}{,} \PYG{k+kt}{size\PYGZus{}t} \PYG{n}{s}\PYG{p}{,} \PYG{n}{arma}\PYG{o}{::}\PYG{n}{mat}\PYG{o}{\PYGZam{}} \PYG{n}{g}\PYG{p}{,} \PYG{k+kt}{size\PYGZus{}t} \PYG{n}{bs}\PYG{p}{)} \PYG{p}{\PYGZob{}}
    \PYG{n}{g}\PYG{p}{.}\PYG{n}{zeros}\PYG{p}{(}\PYG{n}{para}\PYG{p}{.}\PYG{n}{n\PYGZus{}rows}\PYG{p}{,} \PYG{n}{para}\PYG{p}{.}\PYG{n}{n\PYGZus{}cols}\PYG{p}{);}
    \PYG{k}{for} \PYG{p}{(}\PYG{k+kt}{size\PYGZus{}t} \PYG{n}{i} \PYG{o}{=} \PYG{n}{s}\PYG{p}{;} \PYG{n}{i} \PYG{o}{\PYGZlt{}} \PYG{n}{s} \PYG{o}{+} \PYG{n}{bs}\PYG{p}{;} \PYG{n}{i}\PYG{o}{++}\PYG{p}{)}
      \PYG{n}{g}\PYG{p}{(}\PYG{n}{ord}\PYG{p}{[}\PYG{n}{i}\PYG{p}{])} \PYG{o}{+=} \PYG{p}{(}\PYG{l+m+mf}{1.0} \PYG{o}{/} \PYG{n}{bs}\PYG{p}{)} \PYG{o}{*} \PYG{l+m+mi}{2} \PYG{o}{*} \PYG{n}{para}\PYG{p}{(}\PYG{n}{ord}\PYG{p}{[}\PYG{n}{i}\PYG{p}{])} \PYG{o}{+} \PYG{n}{bi}\PYG{p}{(}\PYG{n}{ord}\PYG{p}{[}\PYG{n}{i}\PYG{p}{]);}
  \PYG{p}{\PYGZcb{}}

  \PYG{n}{arma}\PYG{o}{::}\PYG{n}{vec} \PYG{n}{in} \PYG{c+cm}{/* intercepts */}\PYG{p}{,} \PYG{n}{bi} \PYG{c+cm}{/* coeffs */}\PYG{p}{;} \PYG{n}{arma}\PYG{o}{::}\PYG{n}{uvec} \PYG{n}{ord} \PYG{c+cm}{/* function order */}\PYG{p}{;}
\PYG{p}{\PYGZcb{};}
\end{Verbatim}
}
\vspace*{-0.5em}

Note that in this code, we maintain an ordering with the vector {\tt order}; in
other situations, such as training neural networks, we could simply shuffle the columns of the data
matrix in {\tt Shuffle()}.  This objective function will work with any mlpack
optimizer that supports separable or differentiable functions; this includes
all SGD-like optimizers, L-BFGS, simulated annealing, and others.

Next, we wish to define a simple example optimizer that can be used with {\tt
ObjectiveFunction} and other mlpack objective functions.  For this, we must
implement only an {\tt Optimize()} method, and a constructor to set some
parameters.  The code is given below.

\vspace{-0.5em}
%\caption{Simple objective function where each dimension has a parabola with a
%  distinct minimum}
%\vspace{-0.8em}
%\end{listing}
%\vspace{-0.8em}
%\begin{listing}[H]
{\footnotesize
\begin{Verbatim}[commandchars=\\\{\}]
\PYG{k}{struct} \PYG{n}{SimpleOptimizer} \PYG{p}{\PYGZob{}}
  \PYG{n}{SimpleOptimizer}\PYG{p}{(}\PYG{k+kt}{size\PYGZus{}t} \PYG{n}{bs} \PYG{o}{=} \PYG{l+m+mi}{1}\PYG{p}{,} \PYG{k+kt}{double} \PYG{n}{lr} \PYG{o}{=} \PYG{l+m+mf}{0.02}\PYG{p}{)} \PYG{o}{:} \PYG{n}{bs}\PYG{p}{(}\PYG{n}{bs}\PYG{p}{),} \PYG{n}{lr}\PYG{p}{(}\PYG{n}{lr}\PYG{p}{)} \PYG{p}{\PYGZob{}} \PYG{p}{\PYGZcb{}}

  \PYG{k}{template}\PYG{o}{\PYGZlt{}}\PYG{k}{typename} \PYG{n}{FunctionType}\PYG{o}{\PYGZgt{}}
  \PYG{k+kt}{double} \PYG{n}{Optimize}\PYG{p}{(}\PYG{n}{FunctionType}\PYG{o}{\PYGZam{}} \PYG{n}{function}\PYG{p}{,} \PYG{n}{arma}\PYG{o}{::}\PYG{n}{mat}\PYG{o}{\PYGZam{}} \PYG{n}{parameter}\PYG{p}{)} \PYG{p}{\PYGZob{}}
    \PYG{n}{arma}\PYG{o}{::}\PYG{n}{mat} \PYG{n}{gradient}\PYG{p}{;}
    \PYG{k}{for} \PYG{p}{(}\PYG{k+kt}{size\PYGZus{}t} \PYG{n}{i} \PYG{o}{=} \PYG{l+m+mi}{0}\PYG{p}{;} \PYG{n}{i} \PYG{o}{\PYGZlt{}} \PYG{l+m+mi}{5000}\PYG{p}{;} \PYG{n}{i} \PYG{o}{+=} \PYG{n}{bs}\PYG{p}{)} \PYG{p}{\PYGZob{}}
      \PYG{k}{if} \PYG{p}{(}\PYG{n}{i} \PYG{o}{\PYGZpc{}} \PYG{n}{function}\PYG{p}{.}\PYG{n}{NumFunctions}\PYG{p}{()} \PYG{o}{==} \PYG{l+m+mi}{0}\PYG{p}{)} \PYG{p}{\PYGZob{}} \PYG{n}{function}\PYG{p}{.}\PYG{n}{Shuffle}\PYG{p}{();} \PYG{p}{\PYGZcb{}}
      \PYG{n}{function}\PYG{p}{.}\PYG{n}{Gradient}\PYG{p}{(}\PYG{n}{parameter}\PYG{p}{,} \PYG{n}{i} \PYG{o}{\PYGZpc{}} \PYG{n}{function}\PYG{p}{.}\PYG{n}{NumFunctions}\PYG{p}{(),} \PYG{n}{gradient}\PYG{p}{,} \PYG{n}{bs}\PYG{p}{);}
      \PYG{n}{parameter} \PYG{o}{\PYGZhy{}=} \PYG{n}{lr} \PYG{o}{*} \PYG{n}{gradient}\PYG{p}{;}
    \PYG{p}{\PYGZcb{}}
    \PYG{k}{return} \PYG{n}{function}\PYG{p}{.}\PYG{n}{Evaluate}\PYG{p}{(}\PYG{n}{parameter}\PYG{p}{,} \PYG{l+m+mi}{0}\PYG{p}{,} \PYG{n}{function}\PYG{p}{.}\PYG{n}{NumFunctions}\PYG{p}{());}
  \PYG{p}{\PYGZcb{}}
  \PYG{k+kt}{size\PYGZus{}t} \PYG{n}{bs}\PYG{p}{;} \PYG{k+kt}{double} \PYG{n}{lr}\PYG{p}{;} \PYG{c+c1}{// Store batch size and learning rate internally.}
\PYG{p}{\PYGZcb{};}
\end{Verbatim}
}
\vspace{-0.5em}
%\caption{Simple stochastic gradient descent method with a static number of
%iterations to optimize the function parameter ({\tt bs} defines the batch size
%and {\tt lr} the learning rate). )}
%\vspace{-0.6em}
%\end{listing}
%\vspace{-0.8em}

Note that for the sake of brevity we have omitted checks on the batch size
(this optimizer assumes that {\tt function.NumFunctions()} is a multiple of the
batch size) and other typical parts of real implementations, as well as any
static type checking to ensure separability and differentiability.  Still, {\tt
SimpleOptimizer} can work with any mlpack objective function satisfying those
conditions.  This includes mlpack's neural network code, logistic regression,
and other objective functions.

Now, we can find a minimum of {\tt ObjectiveFunction} with {\tt
SimpleOptimizer} using this code:

\vspace*{-0.4em}
{\footnotesize
\begin{Verbatim}[commandchars=\\\{\}]
  \PYG{n}{ObjectiveFunction} \PYG{n}{function}\PYG{p}{;} \PYG{n}{arma}\PYG{o}{::}\PYG{n}{mat} \PYG{n}{parameter}\PYG{p}{(}\PYG{l+s}{\PYGZdq{}0 0 0 0;\PYGZdq{}}\PYG{p}{);}
  \PYG{n}{SimpleOptimizer} \PYG{n}{optimizer}\PYG{p}{;}
  \PYG{n}{std}\PYG{o}{::}\PYG{n}{cout} \PYG{o}{\PYGZlt{}\PYGZlt{}} \PYG{l+s}{\PYGZdq{}objective: \PYGZdq{}} \PYG{o}{\PYGZlt{}\PYGZlt{}} \PYG{n}{optimizer}\PYG{p}{.}\PYG{n}{Optimize}\PYG{p}{(}\PYG{n}{function}\PYG{p}{,} \PYG{n}{parameter}\PYG{p}{);}
\end{Verbatim}
}
\vspace*{-0.4em}

When we run this code, we receive the output: {\tt objective: 123.75 }

The final value of the objective function should be close to the optimal value,
which is the sum of values at the vertices of the parabolas.
This simple example, of course, does not discuss all the intricacies like a more
complex learning rate update routine, but instead presents how the optimization
framework could be used in a simple data science context. Adapting the example
to a real-life application would be straightforward.

An important point to re-emphasize is that the use of templates and
policy-based design allows easy control of behavior by users, simply by
specifying template parameters---or writing custom classes when needed.  In
addition, there is no runtime performance penalty for this flexibility, as
there would be when providing this type of support through inheritance or in
languages such as Python or C.

%Another important component of the optimization framework design is flexibility.
%To this end, we use a C++ programming paradigm known as policy-based design
%\cite{Alexandrescu2001}. In short, this means that the behavior is easily
%controllable by the user, simply by specifying template parameters. For example,
%the SGD class, which is used as basis for other stochastic gradient descent
%methods, accepts a template parameter UpdateType. Thus, if the user wants to
%update the learning rate update strategy that mlpack does not have support for,
%they merely need to implement a optimizer class, and then they can use the type
%SGD<MyCustomLearningRate>. Additionally, because templates are being used, there
%is no additional overhead, as there would be when providing this type of support
%through inheritance or in other languages such as Python or C.

\vspace*{-0.4em}
\section{Statically checking function properties}
\label{sec:static}
\vspace*{-0.3em}

Unfortunately, template metaprogramming can result in some very lengthy error
messages.  Therefore, we must be careful to ensure that a user is able to easily
debug a problem when they implement an objective function or gradient.  To
 improve error message output, we can use C++'s template metaprogramming
support to determine what methods a type has available.  Similarly, we can also
use static compile-time constants to denote the methods that are required by a
specific optimizer.  This is implemented via SFINAE
\cite{Vandevoorde2002sfinae}.  A {\tt static\_assert()} is raised when a given
objective function does not implement the methods required by the optimizer used.

For instance, when attempting to use the L-BFGS optimizer without having a
{\tt Gradient()} function implemented, the user will receive a (comparatively)
simple error message of the form:

\vspace*{-0.5em}
{\footnotesize
\begin{verbatim}
  error: static assertion failed: the FunctionType does not have a correct
      definition of a Gradient() function
\end{verbatim}
}
\vspace*{-0.9em}

\section{Supported optimizers and functions in mlpack}
\vspace*{-0.3em}

Thanks to the easy abstraction, we have been able to provide support for a large
set of diverse optimizers and objective functions.  Below is a list of what is
currently available.

\vspace*{-0.4em}
% I guess to save some space we should group them.
\begin{itemize} \itemsep -1pt
  \item {\bf SGD variants:} Stochastic Gradient Descent (SGD), Stochastic
      Coordinate Descent (SCD), Parallel Stochastic Gradient Descent (Hogwild!),
      Stochastic Gradient Descent with Restarts (SGDR), SMORMS3, AdaGrad,
      AdaDelta, RMSProp, Adam, AdaMax

  \item {\bf Quasi-Newton variants:} Limited-memory BFGS (L-BFGS), incremental
        Quasi-Newton method (IQN), Augmented Lagrangian Method

  \item {\bf Genetic variants:} Conventional Neuro-evolution (CNE), Covariance
        Matrix Adaptation Evolution Strategy (CMA-ES)

  \item {\bf Other:} Conditional Gradient Descent, Frank-Wolfe algorithm, Simulated Annealing

  \item {\bf Objective functions:} Neural Networks, Logistic regression,
      Matrix completion, Neighborhood Components Analysis, Regularized SVD,
      Reinforcement learning, Softmax regression, Sparse autoencoders,
      Sparse SVM
\end{itemize}
\vspace*{-0.4em}

In addition, many methods are currently in development and will be released in
the future.

\vspace*{-0.4em}
\section{Conclusion}
\vspace*{-0.3em}

We have identified that the support for generic and robust optimization is not
currently available in most machine learning toolkits, and acted upon this
observation to provide an easy framework for both implementing new optimizers
and new objective functions to be optimized inside of the mlpack machine
learning library.
The framework provided by mlpack supports a wide array of special cases, and
already has implemented specialized algorithms that outperform their classic
generic alternatives.

\bibliographystyle{plain}
\bibliography{paper}

\end{document}